# Performance Evaluation of Communication Technologies and Network Structure for Smart Grid Applications

Desong Bian [1], Murat Kuzlu [3], Manisa Pipattanasomporn[2], Saifur Rahman[2], Di Shi[1]

[1] GEIRI North America, San Jose, CA
[2] Virginia Polytechnic Institute and State University, Arlington, VA
[3] Old Dominion University, Norfolk, VA

**Abstract:** Design of an effective and reliable communication network supporting smart grid applications requires a selection of appropriate communication technologies and protocols. The objective of this paper is to study and quantify the capabilities of an Advanced Metering Infrastructure (AMI) to support the simultaneous operation of major smart grid functions. These include smart metering, price-induced controls, distribution automation, demand response and electric vehicle charging/discharging applications in terms of throughput and latency. OPNET is used to simulate the performance of selected communication technologies and protocols. Research findings indicate that smart grid applications can operate simultaneously by piggybacking on an existing AMI infrastructure and still achieve their latency requirements.

## 1. Introduction

Nowadays, the electric power grid is transitioning into an intelligent grid, which is called the smart grid [1]. The key to realizing smart grid applications, such as demand response (DR), real-time pricing, automated metering and electric vehicle (EV) related applications, is to appropriately choose corresponding network structures and communication technologies that provide bidirectional end-to-end data communications [2]. Communication networks for smart grid can be presented as a hierarchical multi-layer architecture, which include wide area network (WAN), neighbourhood area network (NAN) and customer premises area network [3]. WAN provides backbones communication for smart grid [4,5]. NAN manages information flow between WANs and customer premises area networks [6]. Customer premise area networks can be further classified as home/building/industrial area network (HAN/BAN/IAN) [7]. They enable communications within customer premises [8].

Popular WAN communication technologies are, such as fiber-optic, powerline communications (PLC), and wireless media using cellular [9]. Popular NAN technologies are, such as ZigBee, Wireless Local Area Network (WLAN), PLC and some long-distance technologies, such as cellular and data over cable services interface specification (DOCSIS) [10]. Various communication technologies, such as ZigBee, WLAN, Z-Wave and PLC, are widely used [11-14]. Fiber-optic communication is one of the fundamental communication technologies for WANs due to its high data rate and immunity to noise [15]. However, it has high upfront investment and maintenance costs [16]. While PLC is a very good candidate for home automation and street light control applications [17,18], its drawbacks are the inability to transmit signals cross a transformer, power line channel distortion, interference, noise, harsh conditions of the power line environment are significant technical issues which affect its implementations [19,20]. ZigBee on the other hand is a cost-effective, low-power, high-efficiency communication technology [21], but interference problems can be a challenge as it shares same channel spectrum with some other protocols [22]. WLAN, well-known as Wi-Fi, is reliable, secure and high-speed. As a result, it is good at supporting short-range communications [23]. However, it is costly and has power consumption as compared with ZigBee and Z-ware [24]. Cellular is one popular radio network, such as 3G and 4G (WiMAX and LTE). WiMAX natively supports the quality of service and real-time two-way broadband communications between nodes [25]. However, WiMAX is expensive and high-power consumption [26]. LTE is a high-speed, low-latency, secure and long-distance wireless communication technology [27]. However, it shares the cellular services with other mobile customers may lead to congestion and reduction the network performance [28]. As summarized above, each type of communication technology has its own advantages and disadvantages. In addition, different smart grid applications have specific communication requirements in terms of their data rate, latency, reliability, coverage range, and security requirements. Hence, it is extremely necessary to conduct performance evaluation of communication technologies for smart grid applications.

As far as the literature review is concerned, there is plentiful research work on performance comparison of communication technologies supporting smart grid applications. In [29], authors compare different communication technologies (i.e., ZigBee, Wi-Fi, Ethernet, etc.) and assess their suitability for deployment to serve smart grid applications, focusing on home automation within a premises area network. In [30], authors propose and analyze the use of LTE multicast between the aggregator and residential (or official) premises for efficient demand response management in smart grids. Effects of communication network performance on dynamic pricing in smart grid are discussed in [31]. In [32], authors provide a comprehensive review of possible communication network infrastructures for metering based on real-world smart grid projects and analyze their advantages/disadvantages in terms of deployment costs, communication range and reliability. Papers [33, 34] propose EV charging management systems comparing between ZigBee and LoRa communication technologies. In [35], authors proposed a communication network model for smart grids considering application requirements, link capacity and traffic settings. Authors in [36,37] propose a heterogeneous communication architecture for smart grids with detailed analysis of communication requirements. However, these work does not take into account practical network infrastructure.

In fact, advanced metering infrastructure (AMI) is one of the most commonly implemented network infrastructures with extremely wide coverage (from WAN to NAN). According to the U.S. Department of Energy's Smart Grid Investment Grant Program (SGIG), majority of the SGIG projects (65 out of 98) are categorized as AMI [38]. Using the existing AMI network to support other smart grid applications besides the metering draws lots of attention recently. These include applications, such as the supervisory control and data acquisition (SCADA) based distribution automation [39], demand side management [40], transmission expansion with phase shifting transformer [41], forced oscillation source locating [42], cooperative control for microgrid [43], transformer identification and phase identification [44], smart energy management [45] and stability analysis for distribution control of microgrid estimation [46]. In addition, authors in [47] discuss a centralized demand curtailment allocation algorithm that can be implemented by piggybacking on AMI. Renewable energy resources can also be monitored and managed via AMI through an hourly DR program [48]. Similarly, smart pricing, smart metering, and optimal EV charging using AMI are introduced in [49, 50, 51]. However, these studies, while focusing on proposing algorithms applications, do not investigate whether an existing AMI network can actually support simultaneous operations of different smart grid applications. In [52], authors discuss technical requirements imposed on the communications network for AMI. Then authors examine each of the AMI application standards found in the open literature based on these requirements. But this paper does not provide simulation, not to mention the analysis for the simultaneous operation of different applications. Authors in [53] carries out an extensive performance evaluation through simulations of current technologies delivering traffic from multiple AMI applications but only focuses on NAN. Authors in [54] discuss scalable distributed communication architectures to support AMI. In [55], a bi-directional communication protocol is introduced considering the effect of AMI environment. The discussion of ZigBee and Power Line Communication (PLC) technologies for AMI is presented in [56-58]. Authors in [59] discuss a heterogeneous WiMAX-WLAN network for AMI communications. A novel path-sharing scheme for an AMI network is presented in [60]. Authors in [61] develop a multipath routing method for AMI networks in a smart grid.

To fully realize benefits of AMI, it is necessary to appropriately choose communication technologies and associated communication networks that provide two-way communications. The comprehensive simulation and analysis of the ability of AMI network to support multiple smart grid applications is still a knowledge gap. To bridge the gap, the objective of this paper is to substantiate the claim that AMI network can support simultaneous operation of other smart grid applications using simulation studies. Considering the extensive literature in this area, the main contributions of this paper are:

- Firstly, popular communication technologies supporting AMI network operation, i.e., fiber-optic, WiMAX, LTE and 900-MHz, are discussed and their performance is simulated in OPNET, commercial software that provides accurate communication simulations [62].
- Secondly, the performance of these communication networks is evaluated considering the simultaneous operation of popular smart grid applications in both NAN and WAN.
- Lastly, the conclusion from this paper provides a comprehensive analysis discussing the ability of AMI to support multiple smart grid applications.

The rest of the paper is organized as follows. Section II summarizes network structures and technologies for smart grid applications. Technical requirements of smart grid applications are summarized in Section III. In Section IV, case studies are discussed, and AMI communication network capability is then evaluated.

## 2. Review of Communication Technologies and Network Structures for Smart Grid Applications

With the rapid transition from a traditional power system into a smart grid, smart metering applications have become widespread. There are a number of AMI rollouts, providing reliable two-way communications between an electric utility and end-use customers. This section provides a comprehensive review of communication technologies deployed in real-world AMI projects in the United States, as well as discusses typical AMI components and communication network structure.

### 2.1. Review of Communication Technologies for AMI

Based on a survey of real-world AMI projects in the U.S. [32], Table I summarizes relevant information of selected AMI projects including their number of smart meters and communication technologies deployed as the backhaul network (in WAN, connecting a control center and base stations) and the smart meter network (in NAN, connecting base stations, data concentrators and smart meters).

From Table 1, it can be seen that fiber optic and WiMAX/LTE are the most popular communication technologies for the AMI backbone network. Between the two choices, the fiber optic option has an advantage over the WiMAX/LTE option in that it can provide higher bandwidth. This is because the bandwidth of a WiMAX/LTE network needs to be shared with other customers in the same cellular network. Furthermore, the fiber optic technology can provide higher reliability level than 4G/LTE during inclement weather conditions.

The 900 MHz RF mesh network appears to be the most popular technology choice to support communications for smart meter networks. This is because it has the good reliability of connection and signal penetration. Also, 900 MHz RF has further reach distance.

### 2.2. Typical AMI Components

Important components that support AMI applications, as well as other major smart grid applications may comprise:

***Control center*** is responsible for supervising overall smart grid operation. For example, it automates data collection process from smart meters; evaluates the quality of the data; generates estimates where errors and gaps exist; and broadcasts the price information or DR event commands.

***Base station*** communicates wirelessly with smart meters and field devices – using fiber optic and connects directly with the control center.

***Data concentrator*** is a combination of software and hardware unit that collects information from smart meters and forwards the information to the utility. Data concentrators are popularly used in densely-populated areas.

***Field devices*** are devices that allow remote control from a central location to accomplish selected smart grid applications, such as distribution automation. Example field devices include remotely controllable voltage regulators, capacitor banks, switches, etc.

***A smart meter*** is a digital meter that can be used to record consumption of electric power/energy and transfer the consumption information to a utility. It can also be used to receive commands or price signals from a utility.

### 2.3. AMI Communication Network Structure

The network as shown in Figure 1 illustrates a possible network structure supporting the AMI application (and perhaps others, such as pricing, EV and distribution automation applications). In this figure, a group of smart meters and field devices are connected to one data concentrator, and then all data concentrators are connected to the control center through the base station. Having data concentrators increases numbers of smart meters and field devices that can be connected to a base station. The communication between a control center and a base station can be fiber optic; that between a base station and data concentrators can be WiMAX/LTE; and that between a data concentrator and smart meters can be RF 900 MHz (per Table 1).

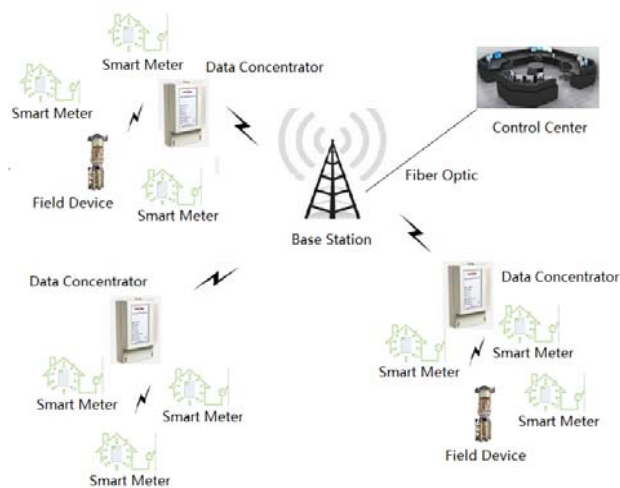

Figure 1. Communication network supporting smart grid applications

## 3. Technical Requirements of Selected Smart Grid Applications

Since different smart grid applications have different characteristics, e.g., data size, data sampling frequency,

TABLE 1. SELECTED REAL-WORLD AMI DEPLOYMENTS

| PROJECT NAME | NUMBER OF METERS | BACKBONE NETWORK | | | | | SMART METER NETWORK | | | | |
|---|---|---|---|---|---|---|---|---|---|---|---|
| | | Fiber | WiMAX/LTE | RF 900 MHz | PLC | Others | WiFi | ZigBee | 900 RF | PLC | WiMAX/LTE |
| Knoxville Smart Grid Community Project, TN | 3,393 | | X | | | | | | X | | |
| Customer Driven Design of Smart Grid Capabilities, WI | 4,355 | | X | | | | | | | | X |
| Advanced Metering Infrastructure Pilot, LA | 4,855 | X | | | | | | | X | | |
| Smart Grid Modernization Initiative, OH | 5,033 | | X | | | | | | X | | |
| Smart Grid Project, IN | 7,474 | X | | | | | | | X | | |
| AMI and Smart Grid Development Program, LA | 10,596 | X | | | | | | | X | | |
| Connected Grid Project, OH | 12,575 | X | | | | | | | X | | |
| Woodruff Electric Advanced Metering Infrastructure Project, AR | 14,450 | | | | X | | | | X | | |
| Leesburg Smart Grid Investment Grant Project, FL | 16,683 | X | | | | | | | | | X |
| Connecticut Municipal Electric Energy Cooperative Project, CT | 23,449 | X | | | | | | | X | | |
| Pacific NW Division Smart Grid Demonstration Project, WA | 30,722 | X | | | X | | | | X | | |
| Smart Grid Team 2020 Program, MD | 38,551 | X | | | | | | | X | | |
| Advanced Metering Infrastructure/Meter Data System, CO | 44,920 | | X | | | | | | X | | |
| Urbank Water and Power Smart Grid Program, CA | 51,928 | | X | | | | | | | | X |
| Smart Grid Program, CA | 52,257 | X | | | | | X | | | | |
| Lafayette Utilities System Smart Grid Project, LA | 63,967 | X | | | | | | X | X | | |
| Advanced Metering Infrastructure/Meter Data System, SD | 68,980 | | X | | | | | | X | | |
| Front Range Smart Grid Cities, CO | 85,328 | X | X | | | | | | X | | |
| AMI Smart Grid Initiative, CA | 85,582 | X | | X | | | | | X | | |
| Smart Grid Initiative, FL | 124,000 | | | X | | | | | X | | |
| Smart Grid Project, NY & NJ | 170,000 | X | | | | | X | X | X | X | X |
| IPC Smart Grid Program, ID | 380,928 | | X | | | | | | | X | |
| Central Main Power (CMP) AMI Project, ME | 622,000 | X | X | | | | | | X | | |
| Smart Currents, MI | 688,717 | | X | | | | | | X | | |
| Smart Grid Initiative, MD | 1.3 million | | X | | | | | | X | X | |
| Smart Grid Project by Centerpoint Energy, TX | 2.1 million | X | | | | | | | X | | |
| Energy Smart Florida, FL | 3 million | X | X | | X | | | | X | | |

latency and reliability requirements, it is, therefore, necessary to ensure proper operation of all smart grid applications especially those sharing the bandwidth with an AMI network. Characteristics of selected smart grid applications, including DR, pricing, metering, EV, Distribution Automation (DA) are summarized in Table 2.

Two types of DR applications are considered: on-demand DR and real-time DR. While on-demand DR schedules a demand reduction at least two hours ahead, real-time DR sends a request to participating customers for a demand reduction in real-time. The pricing application broadcasts time-varying pricing information to end-use customers. Two types of metering applications are considered: on-demand meter reading and meter reading with scheduled time intervals. While on-demand meter reading is used to gather customer meter information as needed, the other kind of meter reading application is to read customer meter data at every fixed time intervals (e.g., 15-minute or an hour). EV application controls the EV charging. DA includes sensing the operating conditions of the distribution grid, and allows making adjustments to improve the overall power flow and distribution-level performance by controlling field devices, such as capacitor banks and switches.

TABLE 2. CHARACTERISTICS OF SELECTED SMART GRID APPLICATIONS [19]

|  | Package Size (bytes) | Data Sampling Frequency (time per day) | Latency (seconds) |
|---|---|---|---|
| **On-demand DR [44]** | 100 | 1 per event | < 60 |
| **Real-time DR** | 100 | As needed | < 5 |
| **Pricing** | 100 | 2-6 | < 60 |
| **On-demand metering** | 100 | As needed | < 15 |
| **Metering with scheduled time intervals** | 1600 - 2400 | 4-6 per residential; 12-24 commercial | < 4 hours |
| **EV Application** | 100 | 2-4 | < 15 |
| **Distribution Automation** | 100 | As needed | < 5 |

The package size shows a number of transmitted/received bytes typically involved in each smart grid application. Data sampling frequency decides the number of packages needed. Latency is the total delay from both algorithm and communication network.

## 4. Case Studies

This section discusses case studies simulated in OPNET to analyze the throughput and latency of different communication options supporting smart grid applications.

### 4.1. CenterPoint Energy - A Reference Smart Grid Project

Based on the AMI deployment reference scenario of the CenterPoint Energy Smart Grid Project [63], the service area (square miles), number of WiMAX towers, data collectors and smart meters are summarized in Table 3.

TABLE 3. DETAIL OF THE CENTERPOINT ENERGY'S AMI PROJECT

|  | Reference Case |
|---|---|
| **Service Area (sq. mile)** | 5,000 |
| **WiMAX Tower** | 112 |
| **Data Collector** | 5,200 |
| **Smart Meter** | 2.2 million |

It can be seen that the density of smart meters in the CenterPoint Energy's service area is 440 meters per sq. mile (2.2 million/5000 sq. mile). The ratio of the WiMAX tower to meter data collectors is 112:5200 or 1:46. The ratio of the data collector to smart meters is 5200:2.2 million or 1:423. And, the ratio of a WiMAX tower to smart meters is 112:2.2 million or 1:19642. These ratios are used to set up the simulation case study as discussed below.

### 4.2. Case Study I: Performance Analysis of the Hybrid Fiber Optic-WiMAX option as the backbone network

#### 4.2.1 Service Area Assumption

The service area of interest covers around 600 sq. miles which is shown in Figure 2. Based on the CenterPoint Energy service area discussed above, it is assumed that 15 WiMAX towers are used to support up to 290,000 smart meters within the service area. The service area of each WiMAX tower is 40 sq. miles. Using Eq. (1), the radius (r) of one WiMAX tower coverage area (hexagonal shape) is calculated to be around 4 miles. In each WiMAX cell, assuming that the ratio of the WiMAX tower to meter data collectors is 1:46 and the ratio of the data collector to smart meters is 1:423, thus there are 46 data concentrators in each WiMAX cell; and each data concentrator is connected with 423 smart meters.

Note that 423 smart meters per data concentrator are used in this case study, which creates the worst case scenario when simulating AMI performance. That is, it can be seen from Table 1 that the density of the smart meters for each data concentrator is much less than 423. Additionally, it is to be noted that smart meter locations within each cell are randomly distributed which is comparable to the real-world environment.

$$Area = \frac{3}{2} * \sqrt{3} * r^2 \qquad (1)$$

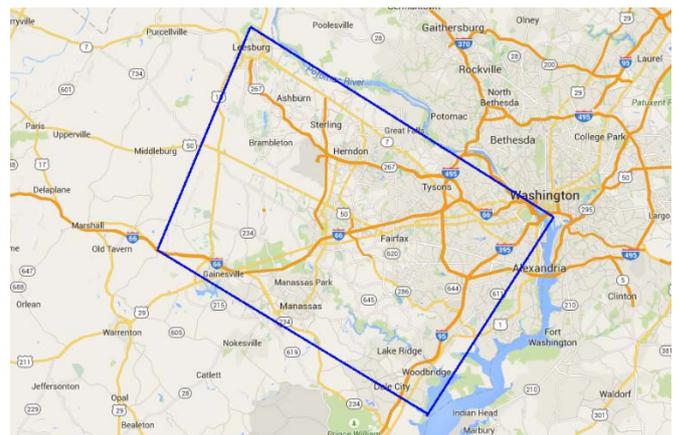

Figure 2. Northern Virginia Service Area

#### 4.2.2 Communication Technology and Network Structure of the Hybrid Fiber-WiMAX Option

The hybrid version of fiber optic-WiMAX technologies is used as a basis to simulate the backbone AMI traffic. That is, fiber optic is selected to serve between the control center and 15 WiMAX base stations; and WiMAX is selected to provide coverage from base stations to data concentrators. The simulation is conducted in OPNET to evaluate the performance of this communication network to support smart grid applications in terms of latency.

To analyze the performance of this network in OPNET, data concentrators are simulated by using subscriber stations (SSs); the BS block is used to simulate the base station; the control center is simulated by using a server station. A detailed case study is simulated in the OPNET with 15 WiMAX towers and 690 data concentrators within the 600 sq. mile area. Figure 3 illustrates how the system is set up in OPNET.

The WiMAX technology used in Case Study I is wireless OFDMA (Orthogonal Frequency Division Multiplexing Access) 20MHz. For this type of WiMAX, the frequency band is 2.3-2.5GHz and the bandwidth is 20MHz. The WiMAX technology provides two-way communications which are Uplink (UL) and Downlink (DL). The UL transfers the information from smart meters to base stations; the DL transfers the information from base stations to smart meters. Both UL and DL are FreeSpace model.

For WiMAX technology, both UL and DL are split into multiple subcarriers with narrow bandwidth. There are four kinds of subcarriers assigned to different functions. Guard subcarrier provides "guard interval" which helps minimize the channel interference. Data subcarriers are used to transfer data. Pilot subcarriers are used for the synchronization. DC (direct current) subcarrier marks the center point of the channel.

The 20MHZ OFDMA WiMAX has 2,048 points of FFT (Fast Fourier Transform) which means it has 2,048 subcarriers in both UL and DL. The detailed classification of subcarriers is summarized in Table 4.

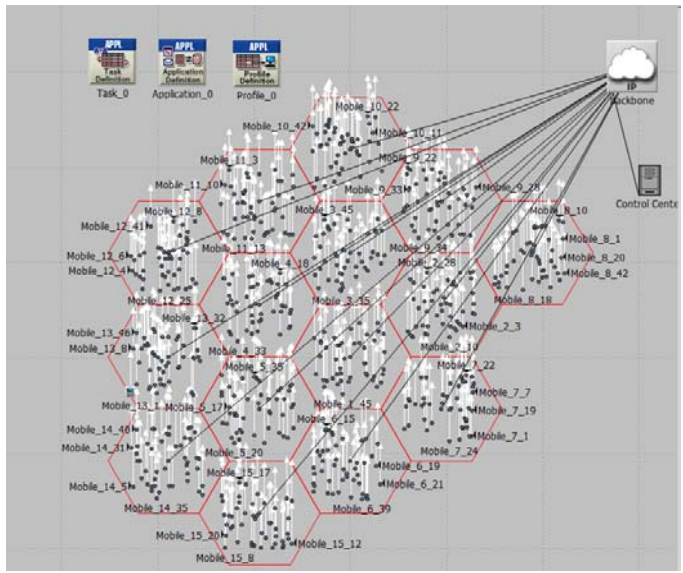

Figure 3. Simulation of Case study I in OPNET

TABLE 4. CLASSIFICATION OF SUBCARRIERS

|  | UL | DL |
|---|---|---|
| **Guard subcarrier from left** | 184 | 184 |
| **Guard subcarrier from right** | 183 | 183 |
| **Data subcarrier** | 1,120 | 1,440 |
| **Pilot subcarrier** | 560 | 240 |
| **DC subcarrier** | 1 | 1 |
| **Total** | 2,048 | 2,048 |

The maximum transmission data rate R that can be achieved in the WiMAX physical layer is defined in the IEEE 802.16 standard as Eq. (2).

$$R = \frac{N_{data} * b_m * c_r}{T_s} \quad (2)$$

where:

$N_{data}$ – the number of data subcarriers;
$b_m$ – the number of bits per modulation symbol (bits);
$c_r$ – Is the coding rate of the modulation (bits/s);
$T_s$ – CP-OFDM symbol time (seconds).

In OPNET, $T_s$ is 100.8 microseconds; $N_{data}$ is 1,120 for UL and 1,440 for DL for the 20 MHz OFDMA. The parameters $b_m$ for QPSK, 16QAM and 64QAM are 2, 4 and 6. In this case, the 64QAM3/4 modulation method is used.

### 4.2.3 Assumptions on Smart Grid Applications

For each application, assumptions on customer participation ratio, the start time of the operation, and the operation duration are summarized in Table 5.

For the real-time DR, metering and pricing applications, the participation ratio of the end-use customer is assumed to be 100%. It means that all end-use customers are involved during the operation of these smart grid applications. It is assumed that there are five field devices located in each cell. For the EV application, it is assumed that half of the end-use customers have electric vehicles and the participation ratio of EV application is thus 50%. For the distribution automation application, only field devices can participate.

TABLE 5. OPERATIONS OF SMART GRID APPLICATIONS

|  | Participation | Operation Begin Time (minute) | | Operation duration (seconds) |
|---|---|---|---|---|
|  |  | 1st Scenario | 2nd Scenario |  |
| **Real-Time DR** | 100% | 50 | 20.8 | 180 |
| **Metering** | 100% | 1, 16, 31, 46 | | 5 |
| **Pricing** | 100% | 21.6 | | 5 |
| **EV** | 50% | 55 | | 5 |
| **DA** | 5 devices in each cell | 45 | | 5 |

In this case study, the simulation lasts for 60 minutes. The metering application's operating frequency is 15 minutes. Thus it operates four times during the simulation interval at the minute 1, 16, 31 and 46. For other four smart grid applications, it is assumed they function only one time during the 60-minute simulation period.

For the real-time DR, its operation duration is assumed to be 3 minutes. For all other smart grid applications, the operation duration is less than 5 seconds.

### 4.2.4 Scenario Description

In both scenarios, five kinds of smart grid applications (real-time DR, metering, pricing, EV application and DA) function in a queue.

In the first scenario, it is assumed that there is no overlap between any two smart grid applications.

In the second scenario, different from the first scenario, there is an overlap in operation between real-time DR and pricing applications.

### 4.2.5 Simulation Results

Simulation results of the first and second scenarios are shown in Figures 4 and 5, respectively. In all case studies, the "seed" which creates the random number generation, is set as 20. As a result, simulation results presented in this paper are average of 20 simulation runs. Since the operation of the real-time DR requires real-time communications, the volume of data exchanging is large. See Figure 4(a) at t=50 and Figure 5(a) at t=20.8. As a result, the latency of this application is a little longer than other smart grid applications.

In scenario one when there is no overlap in operation between any two smart grid applications, the longest latency is around 40ms as shown in Figure 4(b) which is an acceptable latency per the requirement specified in Table 2.

In scenario two, the operation time of the real-time DR application overlaps with that of the pricing application. As a result, the latency of the entire network increases (see Figure 5(b)) to a little longer than 50ms. This is about 10ms increase when running both the real-time DR and pricing applications simultaneously. This implies that an application that sends a 100-byte package to each customer adds about 10ms delay on average to this particular network. Thus, for such applications as meter reading that also sends a 100-byte package another 10ms delay can be expected if it operates together with both real-time DR and pricing applications. For others, such as EV customers which has lower participation and DA which has a limited number of device participation, these applications do not contribute much to added delay due to much lower bandwidth requirements. This implies that if all selected smart grid applications operate simultaneously, the maximum latency will be less than 80ms. This latency is still much lower than the lowest latency requirement of all smart grid applications, i.e., <5 seconds specified in Table 2. Therefore, it can be concluded that all smart grid applications function properly when operating simultaneously.

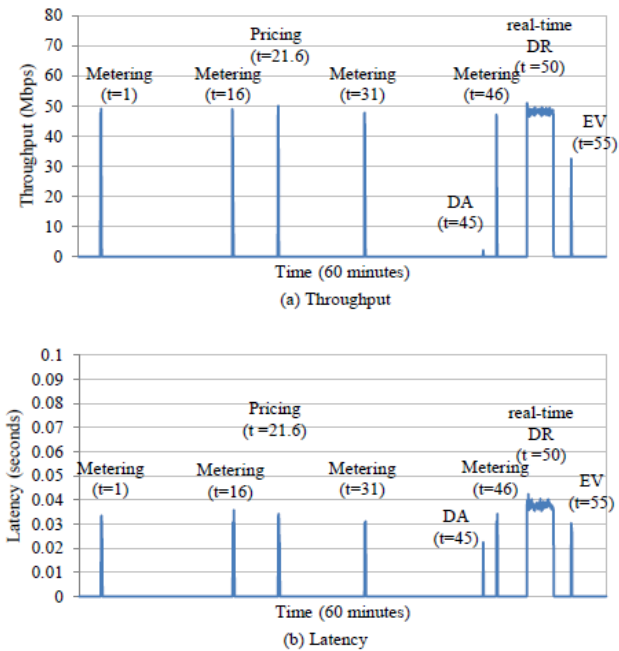

Figure 4. Simulation results: (a) throughput (Mbps) and (b) Latency (seconds) when there is no overlap in operation of different smart grid applications

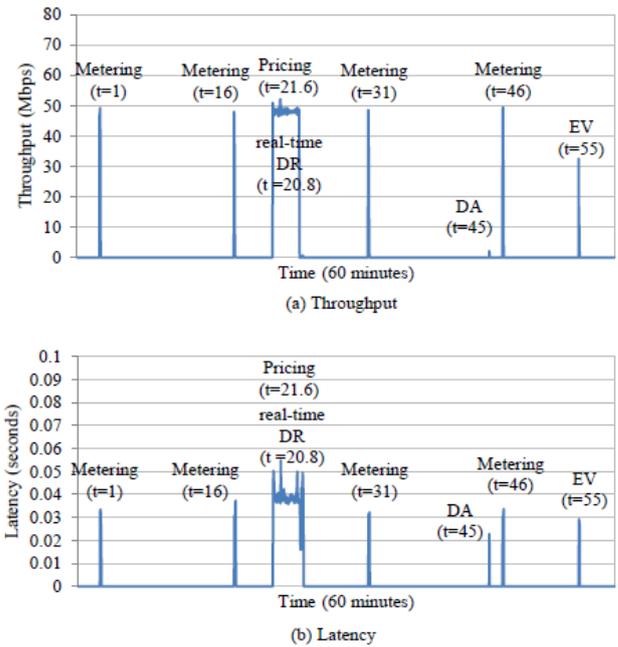

Figure 5. Simulation results: (a) throughput (Mbps) and (b) Latency (seconds) when there is overlap in operation among smart grid applications

### 4.3. Case Study II: Performance Analysis of the Hybrid Fiber Optic-LTE option as the backbone network

#### 4.3.1 Service Area Assumption

In the case study II, same assumptions as the case study I are implemented. Instead of 15 WiMAX base stations used in case study I, 15 LTE base stations are used.

#### 4.3.2 Communication Technology and Network Structure of the Hybrid Fiber-LTE Option

The hybrid version of fiber optic and LTE technologies is used as a basis to simulate the communication traffic between the control center and data concentrators. Fiber optic is selected to serve as between the control center and 15 LTE base stations. LTE is selected to support the smart meter network which covers from base stations to data concentrators.

The simulation is conducted in OPNET to evaluate the performance of smart grid applications in terms of its latency. A detailed case study simulated in the OPNET with 15 LTE towers and 690 data concentrators within the 600 sq. mile area is shown as Figure 6.

The LTE 20 MHz FDD communication technology is applied in case study II. For this LTE technology, the frequency division duplexing (FDD) is used as the duplexing scheme. The LTE technology also provides two-way communications which are Uplink (UL) and Downlink (DL). In this case study, the UL transfers the information from smart meters to base stations; the DL transfers the information from base stations to smart meters. The multipath channel model for LTE's UL is SCFDMA (Single Frequency Division Multiple Access). The LTE frequency band of UL is at 1,920 MHz and the bandwidth of UL is 20 MHz. The multipath channel model for LTE's DL is OFDMA (Orthogonal Frequency Division Multiple Access). The frequency band of DL is at 2,110 MHz and the bandwidth of DL is also 20 MHz. Both UL and DL are FreeSpace model.

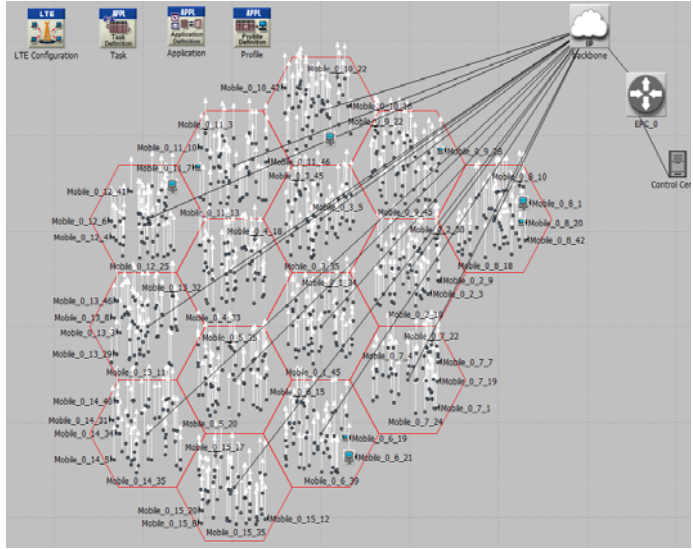

Figure 6. Simulation of Case Study II in OPNET

The modulation type and coding scheme (MCS) index for the LTE applied in this case study is 12 which means the 16-QAM modulation method with ¾ coding rate. The 2*2 MIMO (multiple-input and multiple-output) is applied as the MIMO configuration method for the LTE technology used in this case study. Throughputs of DL and UL can be calculated using Eq. (3).

$$R_{peak} = N_s * \frac{N_m * N_b}{T_s} \quad (3)$$

where:
$N_s$ – the number of data stream;
$N_m$ – the number of modulation symbols per subframe;
$N_b$ – the number of bits per modulations symbol (bits/s);
$T_s$ – the time during of a subframe (second).

In this case study, $N_s$ is 2. Nm is 100 for DL and 50 for UL. $N_b$ is 64 bps. Ts is 71.4 microseconds. And based on the modulation method, the peak data rate provided by OPNET is up to 86.7 and 180 Mbits/s.

*4.3.3 Simulation Results*

Similar to that of case study I, two scenarios are simulated using the fiber optic-LTE network. Simulation results are shown in Figures 7 and 8, respectively. Same as Case Study I, the "seed" is set as 20 and all results shown below are average of 20 simulation results.

Similar to fiber optic and WiMAX cases, around 10ms increased when two applications overlapped. It can be concluded that when all five smart grid applications overlap at the same period, the maximum latency will be less than 80ms. This latency meets the smart grid application requirements specified in Table 2.

*4.4. Performance Analysis of 900 MHz RF to support Smart Grid Applications from Data Concentrator to Smart Meters*

This subsection discusses the latency from a data concentrator to smart meters. Using the data from Section II, each data concentrator is connected with 423 smart meters, and assuming the radius of WiMAX/LTE coverage area is 4 miles. For each customer, the size of data package is 100 bytes which equal to 800 bits. Similarly, two scenarios are considered: non-overlapping and overlapping of operation of any two smart grid applications.

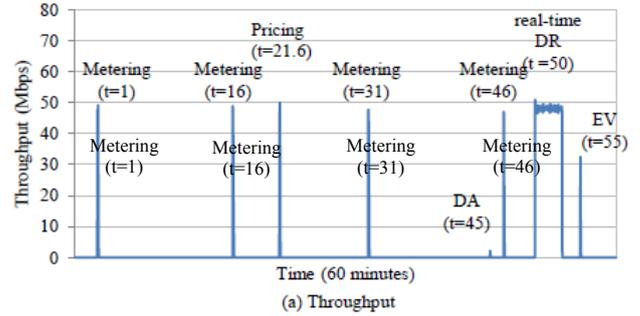

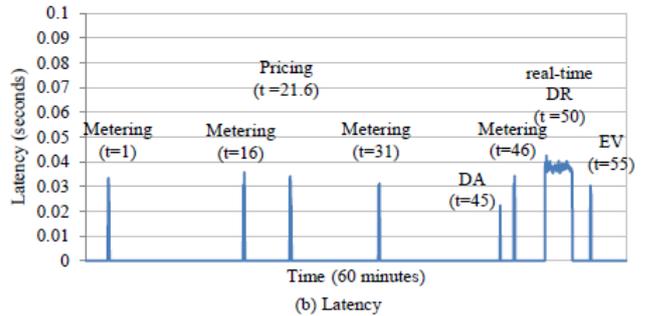

Figure 7. Simulation results: (a) throughput (Mbps) and (b) Latency (seconds) when there is no overlap in operation of different smart grid applications

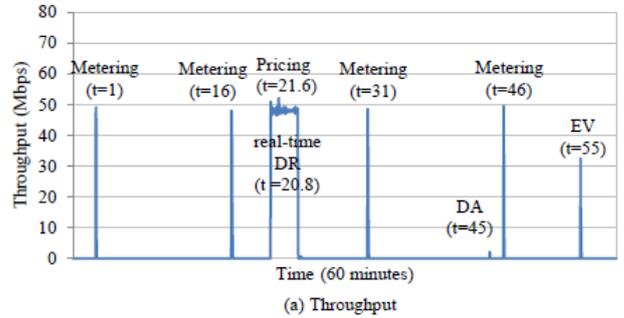

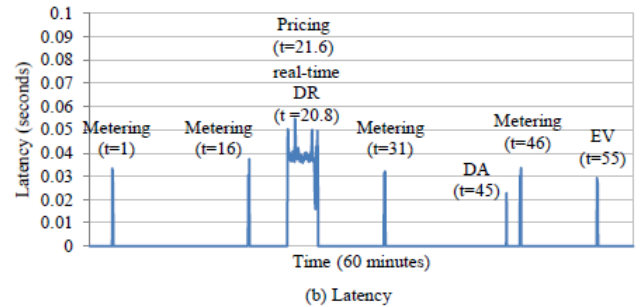

Figure 8. Simulation results: (a) throughput (Mbps) and (b) Latency (seconds) when there is overlap in operation among smart grid applications

For the communication network connecting smart meters and data concentrators, RF technology is widely used in real-world AMI projects. Therefore, the RF mesh network is used as the smart meter network under study. Since RF provides very high reliability, the major consideration of the RF mesh network is its latency.

According to [64], the total delay of an n-hop routed path for one packet sent from one smart meter to the data concentrator can be calculated using Eq. (4).

$$T = \left[n * \left(T_{prop} + \left(\frac{L}{R}\right)\right)\right] + [(n-1) * T_{proc}] \quad (4)$$

where:

$n$ – the number of hops for one packet;
$T_{prop}$ – Propagation delay (second);
$L$ – the length of the packet (bits);
$R$ – the data rate;
$T_{proc}$ – the time spent processing the packet before forwarding it (second).

According to [65, 66], the 900 MHz RF network has the data rate of up to 13.5 Mbps with the coverage of up to 25 miles and allows up to 1000 customers to access. Its coverage and access ability can be implemented to support communications of AMI smart meter network connecting a number of smart meters to a data concentrator. In this section, the package size is 800 bits and the data rate is set as 10 Mbps which is popularly used.

Eq. (5) shows the total latency of an RF network (T), comprising:

$$T = \sum (T_{tran} + T_{prop} + T_{proc}) \quad (5)$$

where:

$T_{tran}$ – delay from pushing the data into a communication channel;
$T_{prop}$ – delay from data traveling from a sender to a receiver;
$T_{proc}$ – delay from collecting data at the receiver.

To calculate the latency, it is assumed that a smart meter and a data concentrator have the same access speed. Eq. (6) shows the calculation of transmission and processing delays.

$$T_{tran} = T_{proc} = \frac{S_p * N_c}{R} \quad (6)$$

where:

$S_p$ – the size of the package (bits);
$N_c$ – the number of customers;
$R$ – the data rate (bps).

To calculate the propagation delay, the distance between each access points and the base station is assumed to be Gaussian distribution. And the propagation speed of signal in free space is as same as the light which is $3*10^8$ m/s. Eq. (7) shows the propagation latency.

$$T_{prop} = \frac{D}{S_{prop}} \quad (7)$$

where:

$D$ – the distance (m);
$S_{prop}$ – the propagation speed (m/s).

*4.4.1 Non-overlapping Scenario*

When there is no overlap operation period between any two smart grid applications, the transmission, propagation and total latency are calculated as shown in Eq. (8), (9) and (10), respectively:

$$T_{tran} = T_{proc} = \frac{S_p * N_c}{R} = \frac{800 * 423}{10 * 10^6 (bps)} = 0.03 \ seconds \quad (8)$$

$$T_{prop} = \frac{D}{S_{prop}} \ll 0.01 \ seconds \quad (9)$$

$$T_{scenario1} = T_{tran} + T_{prop} + T_{proc}$$
$$\leq 0.03 + 0.01 + 0.03 \leq 0.2 \ seconds \quad (10)$$

*4.4.2 Overlapping Scenario*

When there is one overlap operation period between any two smart grid applications, communication traffic throughput is doubled, the latency of the worst case is two times of non-overlapping scenario. The total latency is calculated using Eq. (11).

$$T_{scenario2} = T_{tran} + T_{prop} + T_{proc}$$
$$\leq 0.03 + 0.01 + 0.03 \leq 0.2 \ seconds \quad (11)$$

*4.5. Summary of Case Study Results*

Table 6 summarizes overall case study results on the AMI network latency. As shown, the overall latency of the operation of all five smart grid applications under the overlapping scenario is < 0.06 seconds in the backbone network with fiber-optic WiMAX/LTE; and is < 0.4 seconds in smart meter networks with 900 MHz RF. The overall latency thus meets the latency requirements specified in Table 2. In the real world, the density of smart meters is much less than the assumption used in the case study. The operation frequency for each smart grid application is also smaller than the number used in the case study. As a result, the actual overall latency is expected to be much less than the results shown in Table 6.

TABLE 6. SUMMARY OF CASE STUDY RESULTS

| Latency | Backbone Network (s) | | Smart Meter Network (s) |
|---|---|---|---|
| | Fiber-optic WiMAX | Fiber-optic LTE | 900 MHz RF |
| **Non-overlap** | < 0.05 | < 0.04 | < 0.2 |
| **overlap** | < 0.06 | < 0.05 | < 0.4 |

## 5. Conclusion

With the rapid development of smart grid, there are different aspects of market opportunities and technological applications being deployed simultaneously. For example, there are two demand side management programs that may overlap – one that operates on predefined schedules and the other that operates dynamically based on price. In this paper, the capability of an existing AMI communication network to support multiple types of smart grid applications is evaluated. The observation is that popular communication technologies (i.e., Hybrid fiber optic-WiMAX, Hybrid fiber optic-LTE and 900 MHz RF) implemented with proper communication network structures can support simultaneous operations of programs with predefined schedules and those which operate dynamically based on price.